\documentclass[preprintnumbers, floatfix,letterpaper,aps,prd,epsfig,nofootinbib,
 twocolumn
]{revtex4-1}
\usepackage{bm,graphicx,dcolumn,epstopdf,epsf, latexsym,mathbbol, amssymb,amsmath,color,slashed, mathrsfs,mathcomp, simplewick}
\usepackage[center]{subfigure}
\usepackage{multirow}
\usepackage{makecell}
\usepackage[colorlinks,linkcolor=blue,citecolor=blue,urlcolor=blue]{hyperref}

\begin{document}
\allowdisplaybreaks
 \newcommand{\bq}{\begin{equation}}
 \newcommand{\eq}{\end{equation}}
 \newcommand{\bqn}{\begin{eqnarray}}
 \newcommand{\eqn}{\end{eqnarray}}
 \newcommand{\nb}{\nonumber}
 \newcommand{\lb}{\label}
 \newcommand{\f}{\frac}
 \newcommand{\p}{\partial}
\newcommand{\PRL}{Phys. Rev. Lett.}
\newcommand{\PLB}{Phys. Lett. B}
\newcommand{\PRD}{Phys. Rev. D}
\newcommand{\CQG}{Class. Quantum Grav.}
\newcommand{\JCAP}{J. Cosmol. Astropart. Phys.}
\newcommand{\JHEP}{J. High. Energy. Phys.}
\newcommand{\red}{\textcolor{black}}

\title{Thin Accretion Disk around a four-dimensional Einstein-Gauss-Bonnet Black Hole}

\author{Cheng Liu${}^{a, b}$}
\email{liucheng@zjut.edu.cn}

\author{Tao Zhu${}^{a, b}$}
\email{zhut05@zjut.edu.cn; Corresponding author}

\author{Qiang Wu${}^{a, b}$}
\email{wuq@zjut.edu.cn}

\affiliation{${}^{a}$Institute for Theoretical Physics \& Cosmology, Zhejiang University of Technology, Hangzhou, 310023, China\\
${}^{b}$ United Center for Gravitational Wave Physics (UCGWP),  Zhejiang University of Technology, Hangzhou, 310023, China}

\date{\today}

\begin{abstract}
%

Recently a novel four-dimensional Einstein-Gauss-Bonnet (4EGB) theory of gravity was proposed by Glavan and Lin [D. Glavan and C. Lin, Phys. Rev. Lett. 124, 081301 (2020)] which includes a regularized Gauss-Bonnet term by using the re-scalaring of the Gauss-Bonnet coupling constant $\alpha \to \alpha/(D-4)$ in the limit $D\to 4$. This theory also has been reformulated to a specific class of the Horndeski theory with an additional scalar degree of freedom and to a spatial covariant version with a  Lagrangian multiplier which can eliminate the scalar mode.
 Here we study the physical properties of the electromagnetic radiation emitted from a thin accretion disk around the static spherically symmetric black hole in the 4EGB gravity.  For this purpose, we assume the disk is in a steady-state and in hydrodynamic and thermodynamic equilibrium so that the emitted electromagnetic radiation is a black body spectrum. We study in detail the effects of the Gauss-Bonnet coupling constant $\alpha$ in 4EGB gravity on the energy flux, temperature distribution, and electromagnetic spectrum of the disk. It is shown that with the increases of the parameter $\alpha$, the energy flux, temperature distribution, and electromagnetic spectrum of the accretion disk all increases. Besides, we also show that the accretion efficiency increases as the growth of the parameter $\alpha$. Our results indicate that the thin accretion disk around the static spherically symmetric black hole in the 4EGB gravity is hotter, more luminosity, and more efficient than that around a Schwarzschild black hole with the same mass for a positive $\alpha$, while it is cooler, less luminosity, and less efficient for a negative $\alpha$.

\end{abstract}

\maketitle

\section{Introduction}

Einstein’s theory of general relativity (GR) was proposed over a century ago and has successfully passed a large number of observational tests, mainly in the weak field regime. One of the most impressive results derived from GR is the prediction of black holes. Their existence as physical objects are consistent with available observations of gravitational waves generated due to the merging of black holes by the LIGO experiment \cite{ligo1}, through the extraordinary observation of the M87* black hole shadow by the Event Horizon Collaboration \cite{m87}, and also with the observations of the electromagnetic spectrum emitted from an accretion disk around a black hole \cite{frank, yuan, bambi}. With these observations of black holes in gravitational and electromagnetic spectra, together with their future developments, tests of GR and its alternatives in the strong gravity regime are a hot topic nowadays.

In the strong gravity regime, the observational aspects of black holes are closely related to a narrow region not far from its event horizon, range from the photon sphere to the accretion disk around the black hole. This region is naturally of great significance, as it is perhaps influenced by the possible higher curvature corrections to the Einstein term in GR. The Gauss-Bonnet term and its Lovelock generalization are the most important higher curvature terms studied in various alternative theories beyond GR. However, in four dimensions, the Gauss-Bonnet term is a topological invariant and thus does not contribute to the gravitational dynamics, except it is coupled to a matter field. Recently, a novel 4EGB theory of gravity was proposed by Glavan and Lin \cite{Glavan2020} which includes a regularized Gauss-Bonnet term by using the re-scaling of the Gauss-Bonnet coupling constant $\alpha \to \alpha/(D-4)$ in the limit $D\to 4$. With such scaling, it is shown that the Gauss-Bonnet term can make a non-trivial contribution to the gravitational dynamics in the limit $D \to 4$.  But this modification is in contradiction with the common knowledge and might cause some problems. It is found that this theory is not well defined in the limit $ D \rightarrow 4 $ \cite{W.Y, Gurses, H.Lu, Koba, Henni, Fernan}. Furthermore, the vacua of the model are ill-defined too \cite{shu}. In order to solve these issues, several variants of the original theory have been explored. In \cite{H.Lu}, it is shown that the original 4EGB theory can be reformulated to a specific class of the Horndeski theory with an additional scalar degree of freedom. Its Lovelock generalization as a scalar-tensor theory have also been considered in \cite{Koba}. The similar results have also been explored in \cite{Fernan, Henni} by adding a counter term in $D$-dimensions and then take $D\to 4$ limit. Another regularization procedure is to break the temporal diffeomorphism invariance of the theory \cite{Kat}. In this way, the scalar degree of freedom can be eliminated by a Lagrangian multiplier so the theory have the same number of degrees of freedom as GR.

%

Both the original 4EGB theory and its variants have stimulated a lot of attentions recently. The black hole solutions and their physical properties, such as the shadows \cite{shadow1, shadow2, Wei:2020ght}, quasi-normal modes \cite{shadow1, QNM}, the stability of gravitational perturbations \cite{stability}, the innermost circular orbits of massive and spinning particles \cite{isco1, Zhang:2020qew}, rotating black holes \cite{Kumar:2020owy}, charged black hole in AdS space \cite{Fernandes:2020rpa}, radiating black holes \cite{Ghosh:2020vpc, ghosh, Ghosh:2020vpc}, relativistic stars solution \cite{Doneva:2020ped}, grey-body factor and Hawking radiation \cite{konoplya_greybody_2020, Zhang:2020qam}, stability of the Einstein static universe \cite{li_stability_2020}, gravitational lensing \cite{jin_strong_2020, Islam:2020xmy}, effect of the speed of gravitational waves and the scalar perturbations \cite{JiaXi},  observational constraints on the 4EGB theoretical parameters $ \alpha $ \cite{JiaXi, Timo}, and thermodynamic geometry and phase transitions \cite{Singh:2020xju, Hegde:2020xlv, HosseiniMansoori:2020yfj, Wei:2020poh}, were extensively analyzed. The BTZ black hole in the three-dimensional Einstein-Lovelock gravity has also been explored in \cite{BTZ}.It is worth noting that the holographic implications of the addition of the Gauss-Bonnet term to the anti-de Sitter (AdS) gravity action in four dimensions has been addressed in \cite{Miskovic:2009bm}.

In this paper, we explore the properties of the electromagnetic spectrum emitted from the accretion disk around a static spherically symmetric black hole in 4EGB gravity. For an astrophysical black hole, the study of the electromagnetic spectrum from the accretion process around the black hole is a powerful approach to explore the nature of the black hole spacetime in the regime of strong gravity. This has stimulated a lot works on the studies of the thin accretion disk around various black hole spacetimes, see \cite{Mohad,Faraji2020, Harko2009a, Harko2009, Joshi2014, Kovas2010, Abramowicz2013, Torres, Fish2016, Muller2012, Chow2012, Kovacs2009, Danila, Pun2008, Perez, Gyul2019, Harko2011, Fard2010, Harko2010, Perez2017, Kari2018, Loda, chen2011, chen2012, Liu, spherical, Baner2017, Baner2019, Liu:2020ola} and references therein. Therefore, it is natural to ask whether the Gauss-Bonnet corrections of the 4EGB gravity can appear in the electromagnetic signatures of the accretion disk. To answer this question, we consider a thin relativistic accretion disk model around the 4EGB black hole, which is in a steady-state and in hydrodynamic and thermodynamic equilibrium. In particular, we calculate the energy flux, temperature distribution, and electromagnetic spectrum of the thin accretion disk, and compare them with the standard GR case. The possible effects of the Gauss-Bonnet corrections on the electromagnetic signatures from the thin accretion disk are also explored.

The plan of our paper is as follows. In Sec. II, we present a brief introduction of the recent proposed 4EGB gravity and its static spherically symmetric black hole solution. In Sec. III, we study the geodesic equations for the timelike particles moving in the equatorial plane in the 4EGB black hole. Then in Sec. IV, we study the physical properties of the electromagnetic spectrum emitted from the thin accretion disk around the 4EGB black hole and explore the effects of the Gauss-Bonnet coupling constant $\alpha$ on the energy flux, temperature distribution, electromagnetic spectrum, and the accretion efficiency of the accretion disk. The summary and discussion for this paper is presented in Sec. V.

\section{Black hole solutions in four dimensional Einstein-Gauss-Bonnet gravity}
\renewcommand{\theequation}{2.\arabic{equation}} \setcounter{equation}{0}

In this section, we discuss the theoretical background relevant to the analysis of the thin accretion disk around the 4EGB black hole. 

Let us start with the original action of the $D$-dimensional Einstein-Gauss-Bonnet gravity, which is
\bqn\lb{action}
S_{\rm EGB} = \int d^D x \sqrt{-g} \left(\frac{M_{\rm Pl}^2}{2} R + \alpha R_{\rm GB}^2\right),
\eqn
where $R$ is the Ricci scalar of the spacetime, $R^2_{\text{GB}} \equiv R_{\mu\nu\rho\sigma}R^{\mu\nu\rho\sigma}-4 R_{\mu\nu}R^{\mu\nu}+R^2$ is the Gauss-Bonnet term, and $\alpha$ is the dimensionless coupling constant. Here \red{$M_{\rm Pl}=(8\pi G)^{-1/2}$} is the reduced Planck energy with $G$ being the gravitational constant. In four dimensional spacetime, the Gauss-Bonnet term $R^2_{\text{GB}}$ is a total derivative, therefore it does not contribute to the gravitational dynamics. However,  by re-scaling the coupling constant as $\alpha \to \alpha/(D-4)$,  it is shown by Glaan and Lin \cite{Glavan2020} that the Gauss-Bonnet invariant can make a non-trivial contribution to the gravitational dynamics in the limit $D \to 4$. With such scaling, the action of the four-dimensional Einstein-Gauss-Bonnet gravity (4EGB) in the limit $D\to 4$  can be written as \cite{Glavan2020}
\bqn \lb{4EGB}
S_{\rm 4EGB} = \int d^D x \sqrt{-g} \left(\frac{M_{\rm Pl}^2}{2} R + \frac{\alpha}{D-4} R_{\rm GB}^2\right).
\eqn
Variation of this action with respect to the metric leads to the field equation of 4EGB in the vacuum,
\bqn
R_{\mu \nu} - \frac{1}{2} R g_{\mu\nu} + \frac{\alpha M_{\rm Pl}^{-2}}{D-4} H_{\mu\nu}=0, \lb{field_Eq}
\eqn
where
\bqn
H_{\mu\nu} &=& 2 (RR_{\mu\nu} - 2 R_{\mu \gamma} R^{\gamma}_\nu - 2 R^{\lambda \rho} R_{\mu \lambda \nu \rho} + R_{\mu}^{\lambda \rho \sigma} R_{\nu \lambda \rho \sigma})\nb\\
&& - \frac{1}{2} g_{\mu\nu} R^2_{\text{GB}}.
\eqn
Here we would like to mention that $H_{\mu \nu}$ is proportional to $D-4$ in $D$-dimensional spacetime, therefore in the field equation (\ref{field_Eq}) the Gauss-Bonnet contribution $\alpha H_{\mu \nu}/(D-4)$ can be non-vanishing even in the limit $D\to 4$. 

However, as we mentioned in the previous section, it was found the theory defined in this way has no well-defined limit \cite{W.Y, Gurses, H.Lu, Koba, Henni, Fernan}. There are several schemes to overcome this difficult in the literatures. One approach is to reformulate the original version to a specific class of the Horendeski theory,  which has the following action  \cite{Fernan, Henni, H.Lu, Koba},
\bqn
&&S=\int_{\mathcal{M}}d^4 x\sqrt{-g}\big[ \frac{M_{\rm Pl}^2}{2}R- 2\Lambda_0 + \hat{\alpha}\big(4G^{\mu\nu}\nabla_\mu \phi\nabla_\nu \phi -\nb\\ &&~~~~~\phi \red{R_{\rm GB}^2}+4\square\phi(\nabla \phi)^2+2 (\nabla\phi)^4 \big) \big]+S_m,
\eqn
where $ \phi $ is a scalar field inherent from $ D $ dimensions. In \cite{Fernan, Henni}, it is introduced by a conformal transformation $ g_{ab}\rightarrow e^{2\phi}g_{ab} $ . And in \cite{H.Lu, Koba}, it is introduced by Kaluza-Klein reduction of the metric $ ds^2_D=ds^2_4+e^{2\phi}d\Omega^2_{D-4} $.

Another approach is based on the ADM decomposition analysis and breaks the temporal diffeomorphism invaiance of the theory \cite{Kat}. In this way, the theory only has spatial covariant and one can write the 4-dimensional spacetime metric as
\bqn
ds^2 &=& g_{\mu\nu} dx^\mu d x^\nu\nb\\
&=& -N^2 dt^2+\gamma_{ij}(dx^i+N^i dt)(dx^j+N^j dt),
\eqn
where $N$, $N^i$, and $\gamma_{ij}$ are lapse function, shift vector, and spatial metric respectively. The action of the spatial covariant 4EGB theory can be constructed as \cite{Kat}
\bqn
S_{\text{EGB}}&=& \int dt d^3 x \sqrt{\gamma} N \Bigg\{\frac{M_{\rm Pl}^2}{2} (\;^3R-2 \Lambda+{\cal K}_{ij} {\cal K}^{ij} - {\cal K}^2) \nb\\
&&+\alpha \Big[ -\frac{4}{3} (8 \;^3R_{ij} \;^3R^{ij}-4 \;^3R_{ij} {\cal M}^{ij} - {\cal M}_{ij} {\cal M}^{ij}) \nb\\
&&~~~~ ~ + \frac{1}{2}(8 \;^3 R^2 -4 \;^3 R {\cal M} - {\cal M}^2)\Big]\Bigg\},
\eqn
where 
\bqn
{\cal K}_{ij} &= &\frac{1}{2N} (\partial_t \gamma_{ij} -2 D_{(i}N_{j)} - \gamma_{ij} D^2 \lambda_{\rm GF}),\\
{\cal K} &=& \gamma^{ij} {\cal K}_{ij}, \\
{\cal M}_{ij} &=& \;^3 R_{ij}+ {\cal K} {\cal K}_{ij} - {\cal K}_{ik} {\cal K}^k_j,  \\
{\cal M} &=& \gamma^{ij} {\cal M}_{ij}.
\eqn
with $D$ denoting the spatial covariant derivative and $\lambda_{\rm GF}$ being the  Lagrangian multiplier.

In all the above mentioned theories, the original version or its variants, the static, spherically symmetric black hole solution share the same metric form, 
\bqn
ds^2 &=& - f(r) dt^2 + \frac{dr^2}{f(r)} + r^2 \bar \gamma_{ij} dx^{i} dx^j,\\
f(r) &=& 1+ \frac{r^2 }{16\pi G \alpha} \left(1 \pm \sqrt{1 + \frac{64 \pi \alpha G^2 M}{r^3}}\right),\nb\\
\eqn
where $\bar \gamma_{ij}$ is the metric of $n\equiv D-2$ dimensional unit sphere. In the $D\to 4$ limit, $n \to 2$. $M$ denotes the mass of the black hole and the Gauss-Bonnet coupling constant $\alpha$ is restricted to, \red{$-8 G M^2 \leq 8\pi \alpha \leq G M^2$} \cite{isco1}. The horizon of the black hole is given by
\bqn
r_{\pm} = GM\left(1 \pm \sqrt{1- \red{\frac{8\pi \alpha }{G M^2}}}\right), 
\eqn
where one has two horizons when $\alpha>0$ and one degenerate horizon when \red{$8 \pi \alpha= G M^2$}. When $\alpha$ is negative, the spacetime is also well defined beyond the horizon if \red{$-8G M^2 \leq 8 \pi \alpha \leq 0$}. In this case, the black hole only has one horizon, which is given by
\bqn
r_+ = GM \left(1 + \sqrt{1 -\red{ \frac{8\pi \alpha}{G M^2} }} \right).
\eqn
In this paper, we consider the black hole solution in the region \red{$-8 G M^2 \leq 8 \pi \alpha \leq  G M^2$}. It is worth noting that the above solution was also found in gravity with a conformal anomaly in \cite{conformal}  and was extended to the case with a cosmological constant in \cite{conformal2}.




\section{The Geodesic motion of test particle in 4EGB black hole}
\renewcommand{\theequation}{3.\arabic{equation}} \setcounter{equation}{0}

The accretion disk is formed by particles moving in circular orbits around a compact object, whose physical properties and the electromagnetic radiation characteristics are determined by the space-time geometry around the compact object. For the purpose to study the electromagnetic properties of the thin accretion disk around a 4EGB black hole, let us first consider the evolution of a massive particle in the black hole spacetime. We start with the Lagrangian of the particle,
\bqn
L = \frac{1}{2}g_{\mu \nu} \frac{d x^\mu} {d \lambda } \frac{d x^\nu}{d \lambda},
\eqn
where $\lambda$ denotes the affine parameter of the world line of the particle. For massless particle we have $L=0$ and for massive one $L <0$. Then the generalized momentum $p_\mu$ of the particle can be obtained via
\bqn
p_{\mu} = \frac{\partial L}{\partial \dot x^{\mu}} = g_{\mu\nu} \dot x^\nu,
\eqn
which leads to four equations of motions for a particle with energy $E$ and angular momentum $l$,
\bqn
p_t &=& g_{tt} \dot t  = - \tilde{E},\\
p_\phi &=& g_{\phi \phi} \dot \phi = \tilde{l}, \\
p_r &=& g_{rr} \dot r,\\
p_\theta &=& g_{\theta \theta} \dot \theta.
\eqn
Here a dot denotes the derivative with respect to the affine parameter $\lambda$ of the geodesics. From these expressions we obtain 
\bqn
\dot t = - \frac{ \tilde{E}  }{ g_{tt} } = \frac{\tilde{E}}{f(r)},\\
\dot \phi = \frac{  \tilde{l}}{g_{\phi\phi}} = \frac{\tilde{l}}{r^2 \sin^2\theta}.
\eqn
For timelike geodesics, we have $ g_{\mu \nu} \dot x^\mu \dot x^\nu = -1$. Substituting $\dot t$ and $\dot \phi$ we can get
\bqn
g_{rr} \dot r^2 + g_{\theta \theta} \dot \theta^2 &=& -1 - g_{tt} \dot t^2  - g_{\phi\phi} \dot \phi^2\nb\\
&=& -1 +\frac{\tilde{E}^2}{f(r)}- \frac{\tilde{l}^2}{r^2\sin^2\theta}.
\eqn

We are interested in the evolution of the particle in the equatorial circular orbits. For this reason, we can consider $\theta=\pi/2$ and $\dot \theta=0$ for simplicity. Then the above expression can be simplified into the form
\bqn
\dot r ^2 = \tilde{E}^2 - V_{\rm eff}(r),
\eqn
where $V_{\rm eff}(r)$ denotes the effective potential and is given by
\bqn \lb{Veff}
V_{\rm eff}(r)= \left(1+\frac{\tilde{l}^2}{r^2}\right)f(r).
\eqn

\begin{figure}
	\centering
	\includegraphics[width=3.4in]{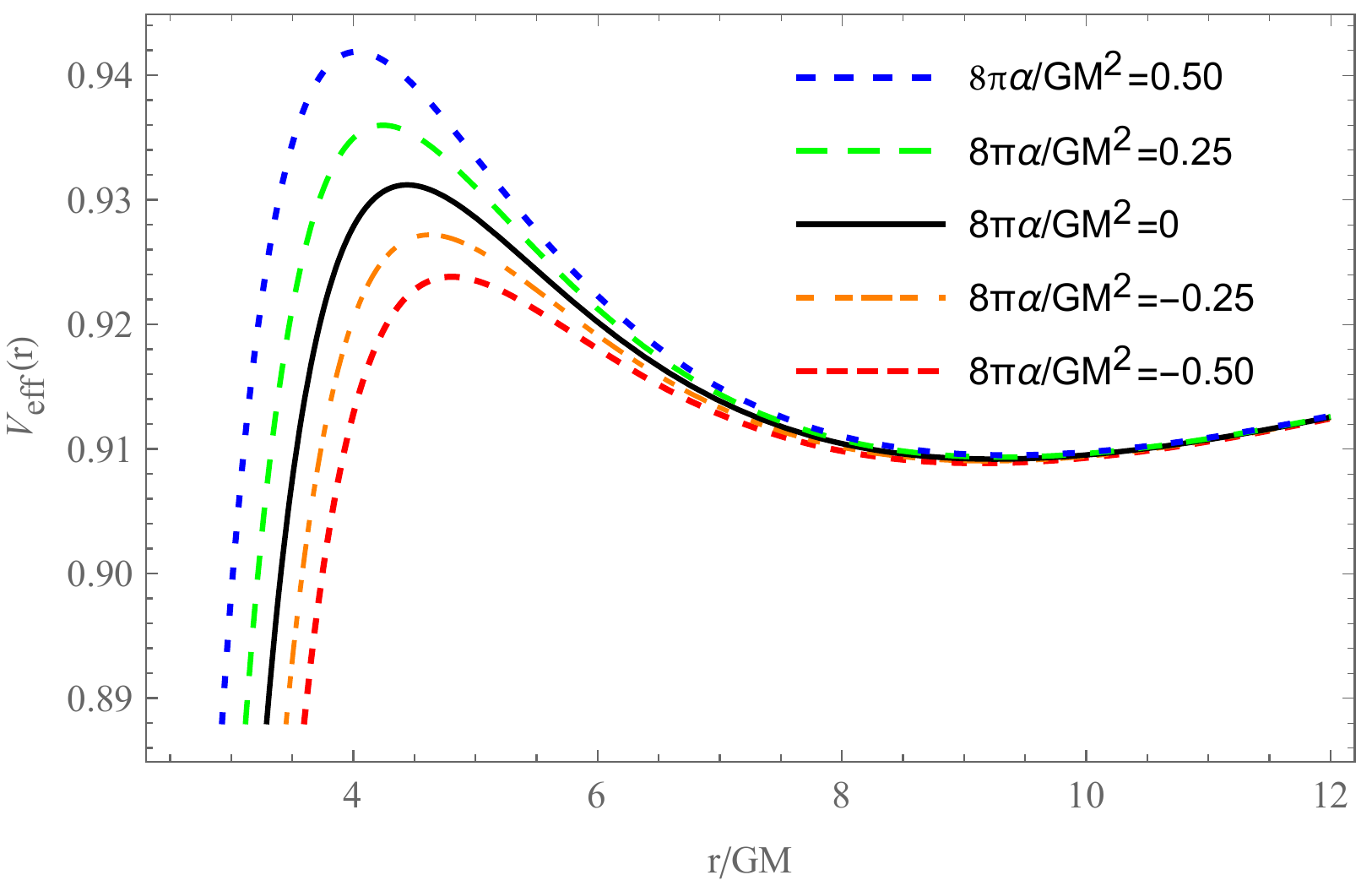}
	\caption{The effective potential of a particle with $\tilde{l}= 3.7$ in the \red{4EGB} black hole spacetime.}
	\label{veff}
\end{figure}
One immediately observes that $V_{\rm eff}(r) \to 1$  as $r \to +\infty$, as expected for an asymptotically flat spacetime. With this case, the particles with energy $\tilde{E} >1$ can escape to infinity, and $\tilde{E} = 1$ is the critical case between bound and unbound orbits. In this sense, the maximum energy for the bound orbits is $\tilde{E}=1$. Fig.~\ref{veff} clearly shows \red{the behaviors how} the effective potential of a particle in the 4EGB black hole spacetime depends on the Gauss-Bonnet coupling $ \alpha$. From this figure, one can easily find that the peak of the effective potential of a particle increases with the coupling constant $\alpha$. The stable circular orbits in the equatorial plane are corresponding to those orbits with constant $r$, i.e., $\dot r^2=0$ and $dV_{\rm eff}(r)/dr=0$. With these conditions, one can write the specific energy $\tilde{E}$, the specific angular momentum $\tilde{l}$, and the angular velocity $\Omega$ of the particle moving in a circular orbit in the 4EGB black hole as 
\bqn
\tilde{E}&=&-\frac{g_{tt}}{\sqrt{-g_{tt}-g_{\phi\phi}\Omega^2}} =\frac{f(r)}{\sqrt{f(r)-\frac{1}{2} r f'(r)}} , \lb{Etilde}\\
\tilde{l}&=&\frac{g_{\phi\phi}\Omega}{\sqrt{-g_{tt}-g_{\phi\phi}\Omega^2}}=\frac{r \sqrt{r  f'(r)}}{\sqrt{2 f(r)-r f'(r)}} ,     \lb{ltilde}\\
\Omega&=&\frac{d\phi}{dt}=\frac{f'(r)}{ \sqrt{2r  f'(r)}}.
\eqn

\begin{figure}
\centering
\includegraphics[width=3.4in]{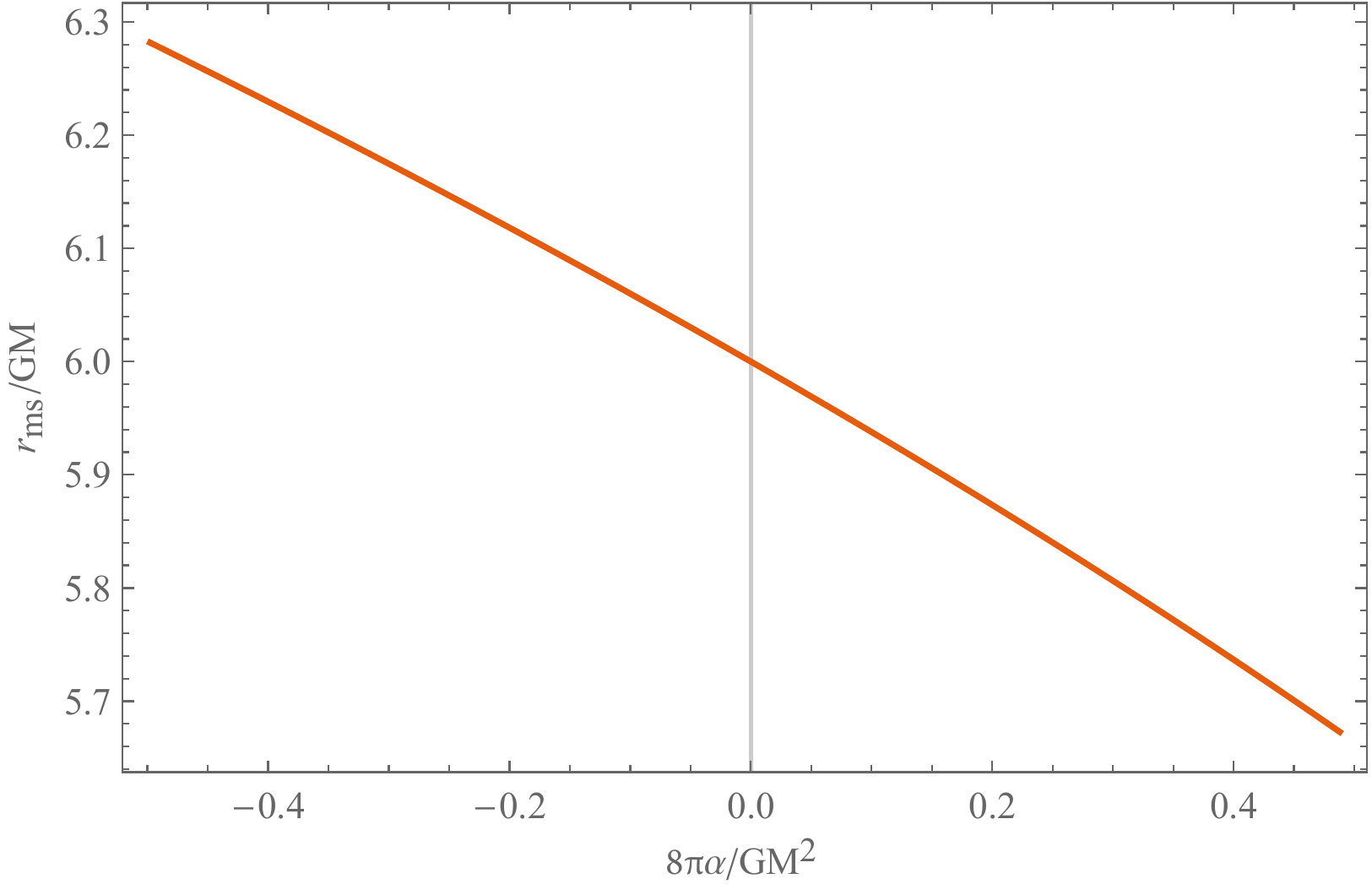}
\caption{The marginally stable orbit radius $r_{\rm ms}$ as a function of the Gauss-Bonnet coupling constant $\alpha$ for particles moving in the thin accretion disk around the 4EGB black hole. }
\label{rms}
\end{figure}

The marginally stable circular orbits around the 4EGB black hole can be determined from the condition 
\bqn
d^2V_{\rm eff}(r)/dr^2 =0.
\eqn
Combining this equation with (\ref{Etilde}) and (\ref{ltilde}) and solving for $r$, the radius of the marginally stable circular orbit can be calculated via
\bqn
r_{\rm ms} = \frac{3 f(r_{\rm ms})f'(r_{\rm ms})}{2f'^2(r_{\rm ms}) - f(r_{\rm ms}) f''(r_{\rm ms})},
\eqn
which does not admit any \red{analytic} solution. We solve it numerically and plot the result in Fig.~\ref{rms}, which shows clearly the radius of the marginally stable circular orbit $r_{\rm ms}$ is decreasing as the increasing of the Gauss-Bonnet coupling constant $\alpha$. In addition, we give an approximate analytic result of the marginally stable circular orbits by making Taylor expansion when $ \alpha $ is small, which yields
\bqn
r_{\text{ms}}=6 GM - \frac{44 \pi }{9M}\alpha + O(\alpha^2).
\eqn
\red{This approximate analytic result clearly shows} that the positive $\alpha$ tends to decrease the radius of the marginally stable circular orbits. Here we would like to mention that the evolution of the massive particles in the 4EGB black hole and the marginally stable circular orbits have also been studied in \cite{isco1}.

\section{The Electromagetic Properties of Thin Accretion Disk Around the 4EGB black hole}
\renewcommand{\theequation}{4.\arabic{equation}} \setcounter{equation}{0}

In this section we consider the steady-state thin accretion disk model and apply it to study the accretion process around the 4EGB black hole. For this purpose we adopt the Novikov-Thorne model of a thin accretion disk consisting of anisotropic fluid moving in the equatorial plane \cite{Novikov1973, Page1974}. In this model, the disk height $H$ is negligible compared to the characteristic radius $R$ of the disk, $H \ll R$. This assumption leads to metric components $g_ {tt} $, $g_ {t\phi } $, $g_ {rr} $, $g_ {\theta \theta}$, and $g_ {\phi \phi}$ only depends on the radial coordinate $r$. The disk is also assumed to be stabilized at hydrodynamic equilibrium, with the pressure and vertical entropy gradient being negligible. An efficient cooling mechanism via heat loss by radiation over the disk surface is assumed to be functioning in the disk, which prevents the disk from collecting the heat generated by stresses and dynamical friction. The thin accretion disk has an inner edge at the marginally stable orbit of the compact object potential, and the accreting matter has a Keplerian motion in higher orbits.

The physical properties of the accretion disk are governed by certain structure equations, which follow from the requirement of the conservation of the rest mass, the energy, and the angular momentum of the fluid. In the thin accretion disk model, the stress-energy tensor of the accreting matter in the disk can be decomposed according to \cite{Novikov1973, Page1974}
\bqn
T^{\mu\nu}=\rho_0 u^\mu u^\nu +2 u^{(\mu}q^{\nu)}+t^{\mu\nu},
\eqn
where
\bqn
u_\mu q^\mu=0, ~~~u_\mu t^{\mu\nu}=0
\eqn
where the quantities $\rho_0$, $q^\mu$ and $t^{\mu\nu}$ represent the rest mass density, the energy flow vector and the stress tensor of the accreting matter, respectively, which is defined in the averaged rest-frame of the orbiting particle with four-velocity $u^\mu$. From the equation if the rest mass is conserved, $\nabla_\mu (\rho_0 u^\mu)=0$, it follows that the time averaged rate of the accretion of the rest mass is independent of the disk radius,
\bqn
\dot{M}_0 &=& -2\pi \sqrt{-g} \Sigma u^r={\rm const.},\\
\Sigma (r) &=& \int^H_{-H}<\rho_0>dz,
\eqn
where $\Sigma(r)$ is the time-averaged surface density and $z$ is the cylindrical coordinates. According to the conservation law of the energy and the law of the angular momentum conservation
\bqn
\nabla_\mu T^{t \mu}=0, ~~~\nabla_\mu T^{\phi \mu}=0,
\eqn
one can obtain the time-averaged radial structure equations of the thin disk around the 4EGB black hole,
\bqn
&&[\dot{M}_0\tilde{E}-2\pi \sqrt{-g} \Omega W^r_\phi]_{,r} =4 \pi r F(r)\tilde{E}, \\
&&~~[\dot{M}_0\tilde{l}-2\pi \sqrt{-g} W^r_\phi]_{,r}=4 \pi r F(r) \tilde{l}.
\eqn
where $W^r_\phi$ is the averaged torque and is given by
\bqn
W^r_\phi=\int^H_{-H}<t^r_\phi>dz, ~~~\sqrt{-{\cal G}}= \sqrt{1+l}r.
\eqn
The quantity $<t^r_\phi>$ is the average value of the $\phi$-$r$ component of the stress tensor over a characteristic time scale $\Delta t$ and the azimuthal angle $\Delta \phi=2 \pi$. By applying the energy-angular momentum relation $\tilde{E}_{,r}=\omega \tilde{l}_{,r}$, the flux $F(r)$ of the radiant energy over the disk can be expressed in terms of the specific energy, angular momentum, and the angular velocity of the orbiting particle in the thin accretion disk around the 4EGB black hole,
\bqn
F(r)=-\frac{\dot{M}_0}{4\pi \sqrt{-g}}\frac{\Omega_{,r}}{(\tilde{E}-\Omega\tilde{l})^2} \int^r_{r_{\rm ms}}(\tilde{E} -\Omega \tilde{l})\tilde{l}_{,r} dr, \lb{energyflux}\nb\\
\eqn
where $r_{\rm ms}$ is the inner edge of the thin accretion disk and is assumed to be at the radius of the marginally stable circular orbit around the 4EGB black hole.

\begin{figure}
\centering
\includegraphics[width=3.4in]{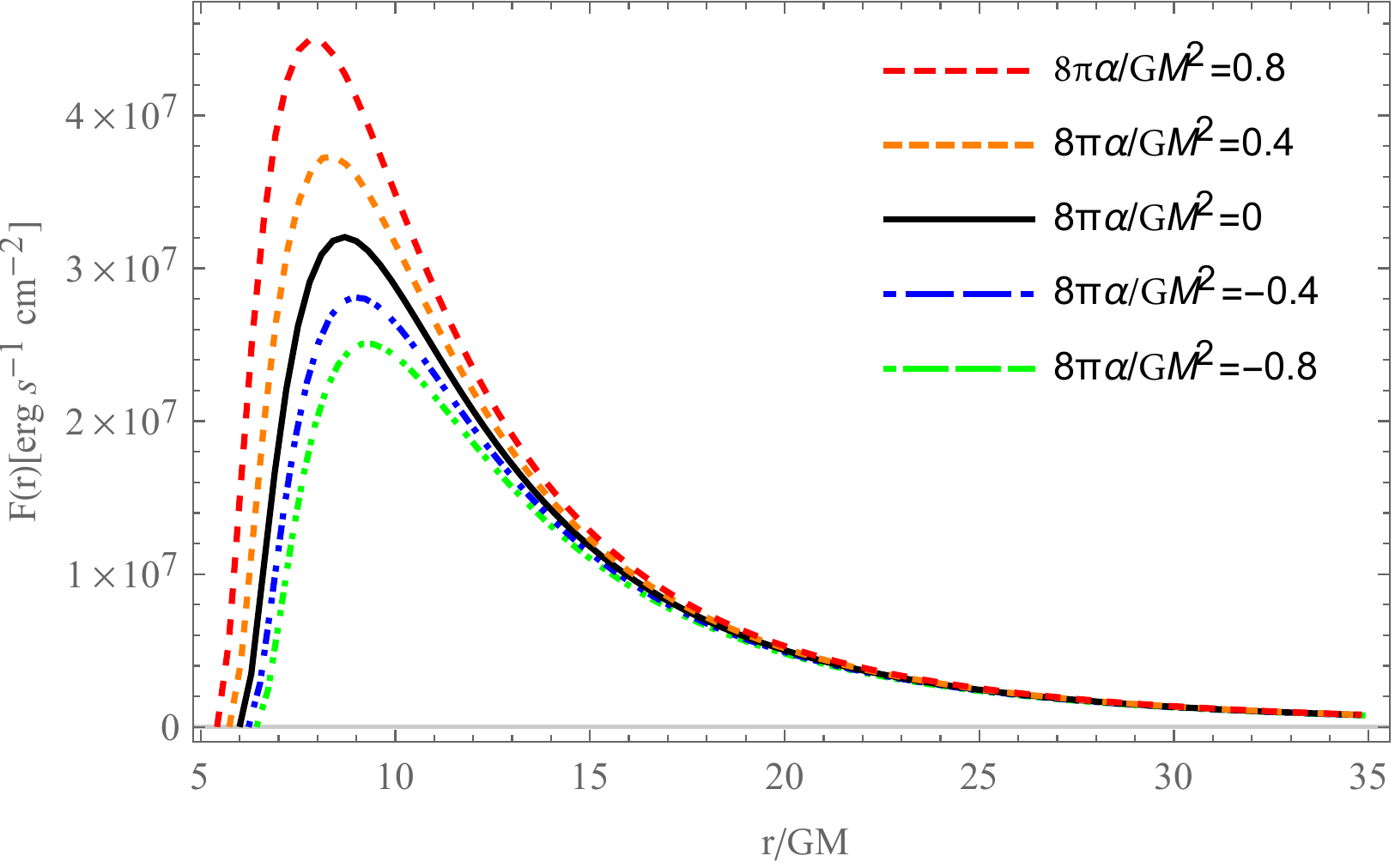}
\caption{Dependence of the radiated energy flux over the thin accretion disk on the radial distance for different values of the Gauss-Bonnet coupling constant $\alpha$. Here the mass of the black hole and the mass accretion rate are set to be $10^{6} M_{\odot}$ and $10^{-12}M_{\odot}/{\rm yr}$ respectively. }
\label{flux}
\end{figure}

We calculate the radiation flux $F(r)$ numerically and illustrate its behavior as a function of the radial distance for different values of the Gauss-Bonnet coupling constant $\alpha$. Following \cite{chen2011, chen2012}, we here consider the mass accretion driven by the 4EGB black hole with a total mass $M=10^6 M_{\odot}$ with a mass accretion rate of $\dot M_0 = 10^{-12} M_{\odot} /{\rm yr}$. In Fig.~\ref{flux} we present the energy flux profile $F(r)$ radiated by a thin accretion disk around the 4EGB black hole for different values of Gauss-Bonnet coupling constant $\alpha$. It is shown that the energy flux grows monotonically with increasing the value of Gauss-Bonnet coupling constant $\alpha$. From the figure, one observes that the energy flux possesses a single maximum, which grows also monotonically with increasing of the value of Gauss-Bonnet coupling constant $\alpha$. At the same time its radial position is shifted towards the location of the horizon. The main reason is that for positive $\alpha$, the effect of the Gauss-Bonnet coupling constant $\alpha$ decreases the radius of the marginally stable orbit so that the lower limit of the integral in (\ref{energyflux}) becomes smaller, while for negative $\alpha$ the radius of the marginally stable orbit increases so that the lower limit becomes larger.

\begin{figure}
\centering
\includegraphics[width=3.4in]{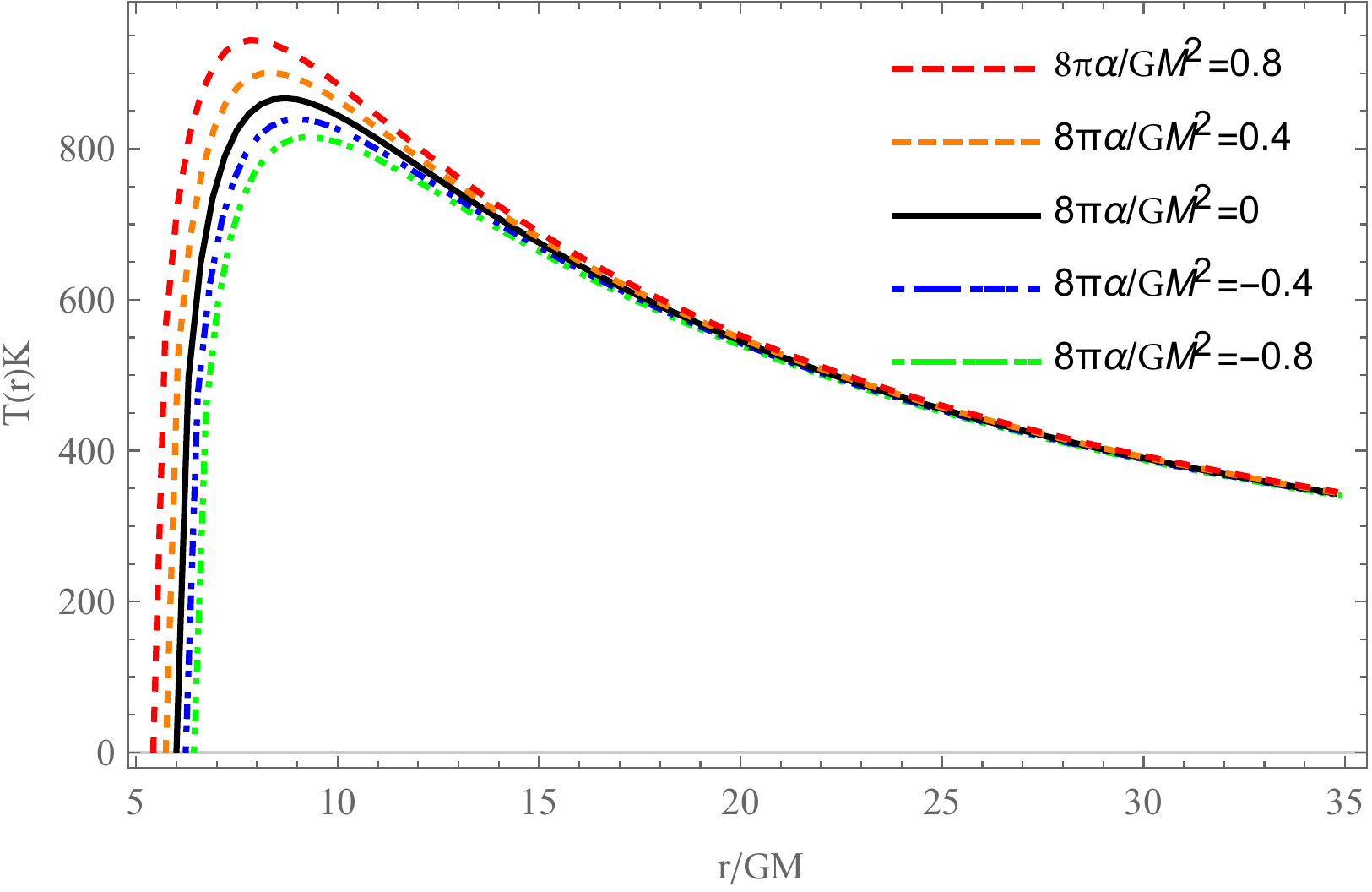}
\caption{The temperature profile of the thin accretion disk around a static spherically symmetric black hole in the 4EGB gravity for different values of the Gauss-Bonent coupling constant $\alpha$. Here the mass of the black hole and the mass accretion rate are set to be $10^{6} M_{\odot}$ and $10^{-12}M_{\odot}/{\rm yr}$ respectively. }
\label{Teff}
\end{figure}

The accreting matter in the steady state thin disk model is supposed to be in thermodynamic equilibrium. The radiation flux $F(r)$ emitted by the thin accretion disk surface will follow Stefan-Boltzmann law. Therefore, the effective temperature of a geometrically thin black-body disk is given by            
\bqn
T_{\rm eff}(r) = \left(\frac{F(r)}{\sigma}\right)^{1/4},
\eqn
where $\sigma = 5.67 \times 10^{-5}\; {\rm erg}\; s^{-1} \;{\rm cm}^{-2}\; K^{-4}$ is the Stefan-Boltzmann constant. In Fig.~\ref{Teff}, we display the radial profile of the effective temperature $T_{\rm eff}(r)$ of the thin accretion disk around the 4EGB black hole for different values of Gauss-Bonnet coupling constant $\alpha$. The figure of the effective temperature shows a similar behavior as that of the energy flux in Fig.~\ref{flux}. It is easy to see from Fig.~\ref{Teff} that the temperature at the fixed radius grows monotonically with increasing the value of the Gauss-Bonnet coupling constant $\alpha$. For a positive value of the Gauss-Bonnet coupling constant $\alpha$, the accretion disk is hotter than that around a Schwarzschild black hole, while it is cooler for a negative $\alpha$.

\begin{figure} 
	\centering
	\includegraphics[width=3.4in]{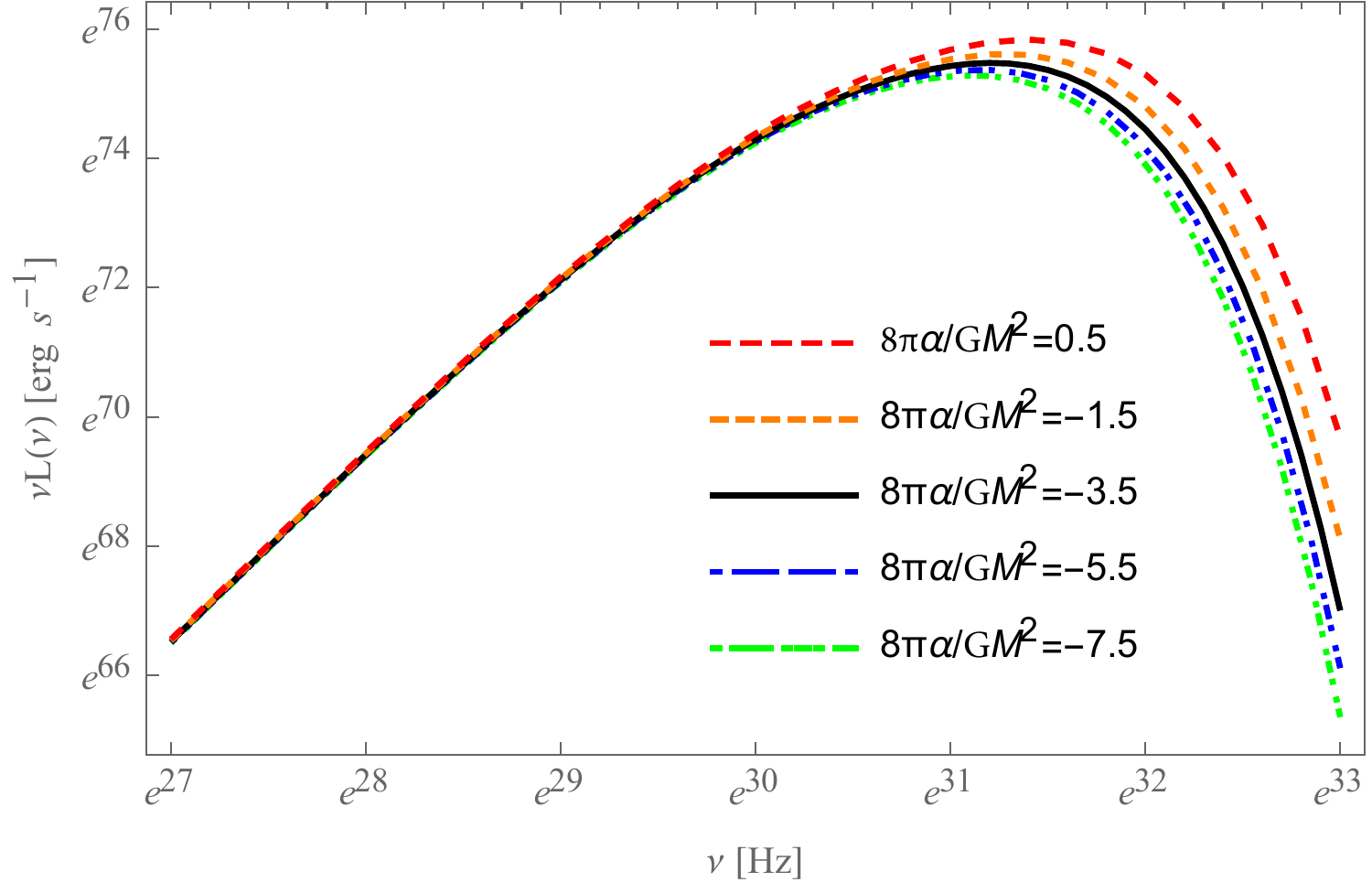}
	\caption{The emission spectrum profile for the thin accretion disk around the 4EGB black hole for different values of the Gauss-Bonnet coupling constant $\alpha$. Here the mass of the black hole and the mass accretion rate are set to be $10^{6} M_{\odot}$ and $10^{-12}M_{\odot}/{\rm yr}$ respectively. }
	\label{Lu}
\end{figure}

Since we consider the radiation emitted by the thin accretion disk surface as a perfect black body radiation, the observed luminosity $L(\nu)$ of the thin accretion disk around the 4EGB black hole has a red-shifted black body spectrum \textcolor{red}{\cite{Cosimo1, Cosimo2}},
\bqn
L(\nu) &=& 4 \pi d^2 I(\nu) \nb\\
&=& \frac{8 \pi h \cos i}{c^2} \int_{r_i}^{r_f} \int_{0}^{2\pi} g^3 \frac{\nu_e^3 r d \phi dr}{\exp{(\frac{h \nu_e}{k_{\rm B} T})}-1},
\eqn
where $i$ is the inclination angle of the thin accretion disk around the 4EGB black hole, $d$ is the distance between the observer and the center of the thin accretion disk, $r_{i}$ and $r_f$ are the inner and outer radii of the disc, $h$ is the Planck constant, $\nu_e$ is the emission frequency in the local rest frame of the emitter, $I(\nu)$ is the Planck distribution, $k_{\rm B}$ is the Boltzmann constant, and $g$ is the redshift factor
\bqn
g=\frac{\nu}{\nu_e}=\frac{k_\mu u^\mu_o}{k_\mu u^\mu_e},
\eqn
where $\nu$ is the radiation fraquency in the local rest frame of the distant observer, $u^\mu_o=(1,0,0,0)$ is the 4-velosity of the observer, and $u^\mu_e=(u^t_e,0,0,\Omega u^t_e)$ is the 4-velosity of the emitter. Since the flux over the disk surface vanishes at $r \to +\infty$ for asymptotically flat geometry, in this paper, we can take $r_i=r_{\rm ms}$ and $r_f = +\infty$. 
To illustrate the effect of the Gauss-Bonnet term in the emission spectrum, we calculate the radiation spectrum $\nu L(\nu)$ numerically and display its behavior as a function of the observed frequency $\nu$ for different values of the Gauss-Bonnet coupling constant $\alpha$ in Fig.~\ref{Lu}. For positive $\alpha$, it is shown that the increasing values of the Gauss-Bonnet coupling constant $\alpha$ produce greater maximal amplitude of the disk emission spectrum as compared to the standard Schwarzschild case, while for negative $\alpha$ it produces a smaller maximal amplitude. From the figure, one also observes that the cut-off frequencies of the emission spectra increases for positive $\alpha$ and decreases for negative $\alpha$, from its value corresponding to the standard Schwarzschild black hole. 

\begin{figure}
\centering
\includegraphics[width=3.4in]{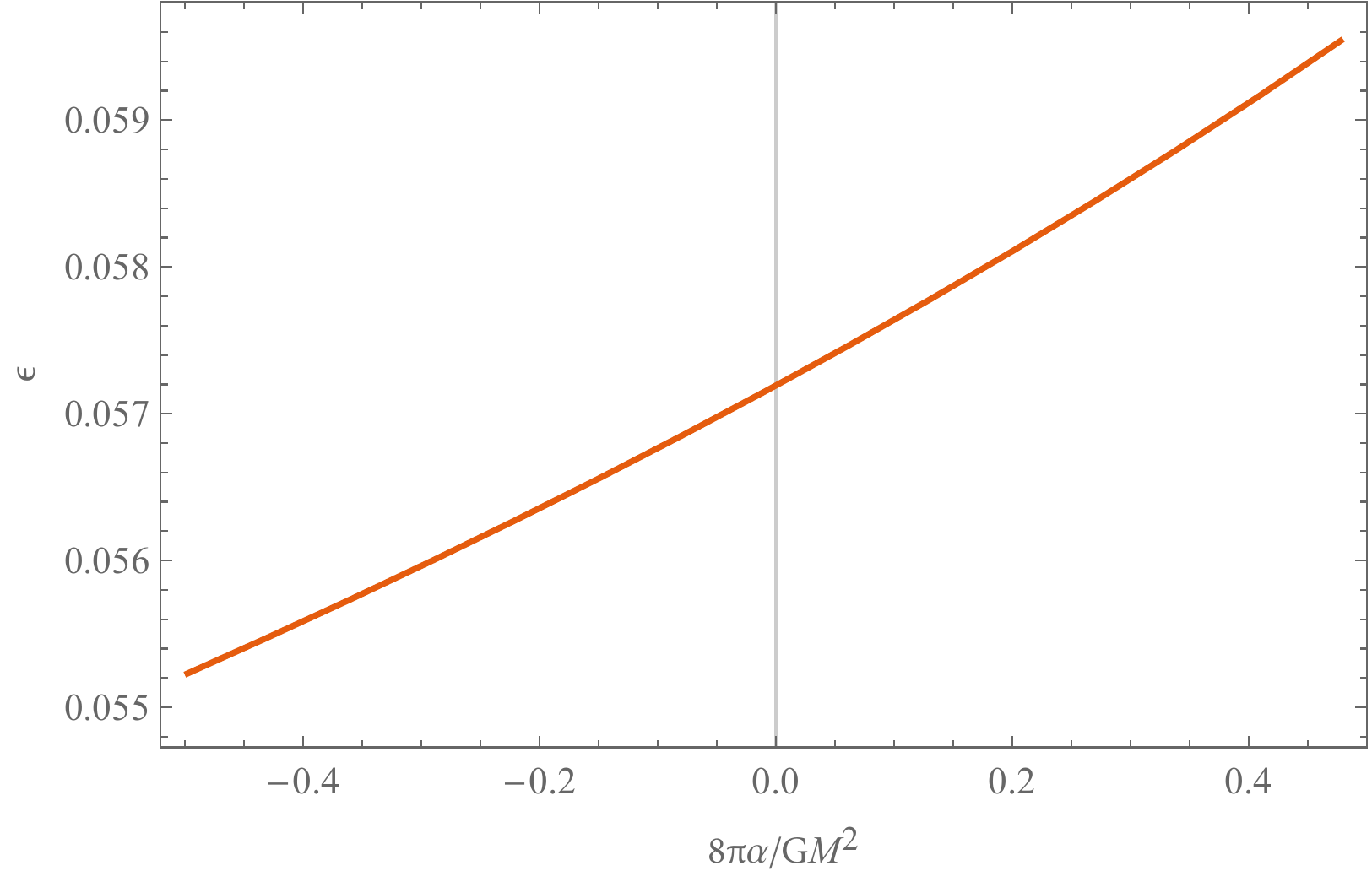}
\caption{The accretion efficiency $\epsilon$ of the 4EGB black hole as a function of the Gauss-Bonnet coupling constant $\alpha$. 
}
\label{epsilon}
\end{figure}

At last, let us consider the accretion efficiency of the 4EGB black hole, which is defined as the ratio of the rate of the radiation of energy of photons escaping from the disk surface to infinity and the rate at which mass-energy is transported to the black hole \cite{Novikov1973, Page1974}. If all the emitted photons can escape to infinity, one can find that the efficiency $\epsilon$ is related to the specific energy of the moving particle in the disk measured at the marginally stable orbit by
\bqn
\epsilon = 1- \tilde E_{\rm ms}.
\eqn
The dependence of the accretion efficiency $\epsilon$ on the Gauss-Bonnet coupling constant $\alpha$ is plotted in Fig.~\ref{epsilon}. It shows that the accretion efficiency $\epsilon$ of the 4EGB black hole increases with the increasing of the values of the Gauss-Bonnet coupling constant $\alpha$. This indicates that the accretion of matter in the 4EGB black hole is more efficient for positive $\alpha$ and lesser efficient for negative $\alpha$ than that in the Schwarzschild black hole. Therefore, the 4EGB black hole with positive Gauss-Bonnet coupling constant $\alpha$ can provide a more efficient engine for transforming the energy of accreting matter into electromagnetic radiation than that with a negative value of $\alpha$.

\section{Conclusion and Discussion}

The 4EGB theory of gravity is a recently proposed theory of gravity, which includes a Gauss-Bonnet curvature corrections to the Einstein term with a coupling constant proportional to $\alpha/(D-4)$ in the limit $D\to 4$. With this formulation, the 4EGB gravity can make a non-trivial contribution to the gravitational dynamics in the limit $D \to 4$. Several variants of this theories have also been explored recently and they have the same static, spherically symmetric black hole solution. In this paper, we study the physical properties of a thin accretion disk around a static spherically symmetric black hole in 4EGB gravity. The physical quantities of the thin accretion disk, such as the energy flux, temperature profile, electromagnetic emission spectrum profiles, and the accretion efficiency have been analyzed in detail for the 4EGB black hole. The effects of the Gauss-Bonnet coupling constant $\alpha$ on these physical quantities have been explicitly obtained. It is shown that with the increases of the parameter $\alpha$, energy flux, temperature distribution, and electromagnetic spectrum of the disk all increases. The main reason for this kind of behaviors is that for positive $\alpha$, the effect of the Gauss-Bonnet coupling constant $\alpha$ decreases the radius of the marginally stable orbit so that the lower limit of the integral in (\ref{energyflux}) becomes smaller, while for negative $\alpha$ the radius of the marginally stable orbit increases so that the lower limit becomes larger. In addition, we also show that the accretion efficiency increases as the growth of the parameter $\alpha$. Our results indicate that the thin accretion disk around the static spherically symmetric black hole in the 4EGB gravity is hotter, more luminosity, and more efficient than that around a Schwarzschild black hole with the same mass for a positive $\alpha$, while it is cooler, less luminosity, and less efficient for a negative $\alpha$.

With the above main results, we would like to mention two directions that can be carried out to extend our analysis. First, in this paper, we only focus on the static and spherically symmetric black hole case. It is very interesting to explore behaviors of the electromagnetic emission spectrum profiles of the thin accretion disk around a rotating black hole in the 4EGB theories. However, the construction of the rotating solution is not an easy task and currently the rigorous rotating black hole via solving the field equation has not been found yet. Although there is an effective rotating solution generated from the static and spherically symmetric black hole by using the Newman-Janis procedure \cite{Wei:2020ght, Kumar:2020owy}, it is still not clear whether this solution is an exact one that satisfies the field equation of the 4EGB theories. In order to explore the features of the electromagnetic emission spectrum with rotation effects, we prefer to find the rotating solution by solving the field equation and this will be considered in our future work. 

Second, when the rotating solution is found, it is also interesting to constrain the black hole parameters, including the angular momentum and the Gauss-Bonnet parameter $\alpha$, by using the observation spectra of the X-ray binaries. For example, by using the continuum-fitting method \cite{Cosimo1}, one is able to measure the angular momentum of the stellar-mass black holes and constrain the deviations from Kerr by using the X-ray data from black hole binaries. At present, the observational constraints of the 4EGB theory mainly come from Refs. \cite{JiaXi} and \cite{Timo}. In \cite{Timo}, the authors have considered the constraints on $\alpha$ from various physical systems, including the solar system experiments, binary pulsars, and cosmological observations. It is shown that  the tightest constraints come from observations of the periapsis advance of the LAGEOS II satellite and from the observation of binary black hole systems, which both lead to $ |16 \pi G \alpha|\lesssim 10^{10} m^2$. In \cite{JiaXi}, the authors obtain the constraints, $ -7.78\times 10^{-16} \lesssim 16 \pi G \alpha H^2 \lesssim 3.33\times 10^{-15} $ with $H$ being the Hubble parameter from the current observation of the speed of GWs measured by GW170817 and gamma ray burst event GRB 170817A. \red{Compared} to these existing constraints, the X-ray binaries can provide a very different environment for testing 4EGB theory and may be more sensitive to the strong gravity behaviors of the 4EGB theory. We expect to return to the above issues soon in future studies to possibly extend some of these results.

\section*{ACKNOWLEDGEMENTS}

This work is supported by National Natural Science Foundation of China with the Grants No.11675143, the Zhejiang  Provincial  Natural Science Foundation of China under Grant No. LY20A050002, and the Fundamental Research Funds for the Provincial Universities of Zhejiang in China under Grants No. RF-A2019015.

\end{document}